\documentclass[useAMS,usenatbib]{mn2e}
\usepackage[letterpaper,total={17.8cm,24.0cm},centering]{geometry}

\usepackage{graphicx}
\usepackage{subfigure}
\usepackage{deluxetable}
\usepackage{verbatim}
\usepackage{natbib}
\usepackage{amsmath,amssymb}
\usepackage{color}
\usepackage[draft]{hyperref}
\newcommand{\mytilde}{\textasciitilde}

\title[Disk galaxy morphology and the star-forming main sequence]{Galaxy Zoo: the dependence of the star formation-stellar mass relation on spiral disk morphology}
\author[Willett et al.]{
  \parbox[t]{16cm}{
  Kyle W. Willett$^{1}$\thanks{E-mail: willett@physics.umn.edu},
  Kevin Schawinski$^{2}$,
  Brooke D. Simmons$^{3}$,
  Karen L. Masters$^{4,5}$,
  Ramin A. Skibba$^{6}$,
  Sugata Kaviraj$^{3,7}$,
  Thomas Melvin$^{4}$,
  O. Ivy Wong$^{8}$,
  Robert C. Nichol$^{4,5}$,
  Edmond Cheung$^{9,10}$,
  Chris J. Lintott$^{3}$,
  Lucy Fortson$^{1}$
  \\
  }\\
$^{1}$School of Physics and Astronomy, University of Minnesota, 116 Church St SE, Minneapolis, MN 55455, USA \\
$^{2}$Institute for Astronomy, Department of Physics, ETH Z\"urich, Wolfgang-Pauli-Strasse 16, CH-8093 Z\"urich, Switzerland \\
$^{3}$Oxford Astrophysics, Denys Wilkinson Building, Keble Road, Oxford OX1 3RH, UK \\
$^{4}$Institute of Cosmology and Gravitation, University of Portsmouth, Dennis Sciama Building, Portsmouth PO1 3FX, UK \\
$^{5}$SEPnet, South East Physics Network, UK \\
$^{6}$Center for Astrophysics and Space Sciences, University of California San Diego, 9500 Gilman Dr, San Diego, CA 92093, USA \\
$^{7}$Centre for Astrophysics Research, University of Hertfordshire, College Lane, Hatfield, Herts, AL10 9AB, UK \\
$^{8}$International Centre for Radio Astronomy Research, University of Western Australia, 35 Stirling Hwy, Crawley, WA 6009, Australia \\
$^{9}$Department of Astronomy and Astrophysics, 1156 High Street, University of California, Santa Cruz, CA 95064, USA\\
$^{10}$Kavli IPMU (WPI), The University of Tokyo, Kashiwa, Chiba 277-8583, Japan\\
}

\begin{document}
\date{Accepted 11 Feb 2015}
\pagerange{\pageref{firstpage}--\pageref{lastpage}} \pubyear{2015}
\maketitle
\label{firstpage}


\begin{abstract}
We measure the stellar mass-star formation rate relation in star-forming disk galaxies at $z\le0.085$, using Galaxy~Zoo morphologies to examine different populations of spirals as classified by their kiloparsec-scale structure. We examine the number of spiral arms, their relative pitch angle, and the presence of a galactic bar in the disk, and show that both the slope and dispersion of the $M_\star$-SFR relation is constant when varying all the above parameters. We also show that mergers (both major and minor), which represent the strongest conditions for increases in star formation at a constant mass, only boost the SFR above the main relation by $\sim0.3$~dex; this is significantly smaller than the increase seen in merging systems at $z>1$. Of the galaxies lying significantly above the $M_\star$-SFR relation in the local Universe, more than $50\%$ are mergers. We interpret this as evidence that the spiral arms, which are imperfect reflections of the galaxy's current gravitational potential, are either fully independent of the various quenching mechanisms or are completely overwhelmed by the combination of outflows and feedback. The arrangement of the star formation can be changed, but the system \emph{as a whole} regulates itself even in the presence of strong dynamical forcing. 
\end{abstract}



\section{Introduction} \label{sec-intro}

Observations at a range of redshifts have established that the star formation rate (SFR) of a galaxy is strongly correlated to its stellar mass ($M_\star$). This ``star-forming main sequence'' (SFMS) is nearly linear and has remarkably small scatter at low redshifts \citep{bri04,sal07}. Recent observations of star-forming galaxies at high redshifts show that this main sequence remains out to high redshift, but the normalisation shifts upward so that galaxies of the same $M_\star$ have a higher SFR at high redshift \citep{noe07,dad07}. The main sequence has been interpreted by \citet{bou10} and \citet{lil13} as the result of the balancing of inflows of cosmological gas and outflows due the feedback. Galaxies self-regulate to remain in a state of homeostasis as they convert baryons from gas to stars. This relation is found in all models where the star formation history of star-forming galaxies is relatively flat over cosmic time, and is insensitive to the details of the feedback mechanism \citep{hop14}. 

As star-forming galaxies may exhibit a wide range of physical appearances in optical images, the natural question can be asked whether the specifics of this morphology and its underlying dynamical processes have any effect on this homeostasis and therefore the galaxy's location relative to the SFMS. If the details of a galaxy's physical appearance are correlated with position relative to the main sequence, then the dynamical processes that give rise to them -- such as bar formation and spiral arm pitch angle -- are a fundamental aspect of the galaxy's regulatory mechanism. If, on the other hand, these features are not correlated, then there are two options: either galaxy substructure is simply not relevant to the overall $M_\star$-SFR relationship, or the regulatory mechanism overcomes the local effect of substructure in all circumstances. This ultimately relates to the physical processes that control the overall strength of the regulator in each galaxy.

The fact that star-forming galaxies live on the SFMS is one of the key observations that has been driving the development of new descriptions of how galaxies evolve \citep[eg, ][]{sch07b}. \citet{pen10a,pen12} argue that galaxies grow in stellar mass during their life as star-forming galaxies on the main sequence before having their star formation quenched either by an external mechanism (`environment quenching') or by an internal mechanism (`mass quenching').  However, \citet{del12} point out that, because of a `history bias,' galaxies of identical stellar mass may have different environmental histories that are difficult to disentangle, thus making the mass versus environment debate inherently ill-posed \citep[see also][]{van08b}. In addition, Galaxy~Zoo data has shown that the environmental dependence of galaxy star formation and that of morphology are not equivalent, mainly because of the abundance of quenched spiral galaxies, a large fraction of which are satellite galaxies \citep{ski09,bam09}. In any case, life on the main sequence appears to be governed by the action of the regulator to balance gas inflows and outflows \citep{lil13}, making the SFMS a central process in galaxy evolution.

In this paper, we use data from the Sloan Digital Sky Survey \citep[SDSS;][]{yor00,str02,aba09} in combination with Galaxy~Zoo, the largest database of visual classifications of galaxy structure and morphology ever assembled \citep{lin08,lin11,wil13}, to test whether disk structure affects a galaxy's star formation properties. We use the WMAP9 cosmology parameters of $(\Omega_m,\Omega_\Lambda,h)=(0.258,0.718,0.697)$ \citep{hin13}.


\section{Data} \label{sec-data}

Photometric and spectroscopic data for all galaxies in this paper comes from optical observations in the SDSS~DR7. The morphological data is drawn from citizen science classifications in Galaxy~Zoo. Detailed classifications of disk morphologies, including arm pitch angle, number of spiral arms, and presence of a galactic bar, are taken from the Galaxy~Zoo~2 (GZ2) catalog \citep{wil13}. Merging pairs of galaxies are taken from the catalog of \citet{dar10a}, all of which lie in the redshift range $0.005<z<0.1$. Post-merger spheroidal galaxies without an obvious, separated companion are specifically excluded from our sample.

Stellar masses and star formation rates are computed from optical diagnostics and taken from the MPA-JHU catalogue \citep{kau03a,bri04,sal07}. We use updated masses and activity classifications from the DR7 database.\footnote{\url{http://home.strw.leidenuniv.nl/\mytilde jarle/SDSS/}} We select only galaxies with $M_\star > 10^8 M_\odot$, for which \citet{bri04} estimate that the MPA-JHU sample is complete. Star-forming galaxies are separated from other emission-line galaxies using the standard BPT classification \citep*{bal81} below the \citet{kau03} demarcation. Galaxies classified as star-forming but with low signal-to-noise ratio $(S/N > 3)$ are also excluded. Both $M_\star$ and SFR are measured using median values extracted from the probability distribution functions.

The spiral nature of the star-forming disk galaxies is identified according to the following thresholds in the GZ2 spectroscopic sample, where $p$ is the debiased vote fraction and $N$ the weighted number of total votes: $p_\textrm{features/disk} > 0.430$, $p_\textrm{not~edgeon} > 0.715$, $p_\textrm{spiral}>0.619$, and $N_\textrm{spiral}>20$. These cuts are chosen to ensure reliable identification and with enough data points such that spiral substructure has a reasonable estimate of the associated uncertainty. Sub-classes of spiral structure are identified by weighting each galaxy according to the fraction of votes received in each morphological category. 

\begin{table} 
 \begin{tabular}{@{}lrcrcl}
 \hline
\multicolumn{1}{c}{Sample} &
\multicolumn{1}{c}{$N$} &
\multicolumn{1}{c}{$\alpha$} &
\multicolumn{1}{c}{$\beta$} &
\multicolumn{1}{c}{$\sigma_\alpha$} &
\multicolumn{1}{c}{$\sigma_\beta$} 
\\ 
\hline
\hline						
SF galaxies  & 48405  & $0.72$  & $-7.08$  &  $7.13\times10^{-6}$  & $6.98\times10^{-4}$  \\
\hline                                                                                   
1 arm        & 288    & $0.74$  & $-7.18$  &  $2.70\times10^{-5}$  & $2.69\times10^{-3}$  \\
2 arms       & 5635   & $0.78$  & $-7.69$  &  $3.70\times10^{-5}$  & $3.78\times10^{-3}$  \\
3 arms       & 995    & $0.71$  & $-6.99$  &  $4.49\times10^{-5}$  & $4.65\times10^{-3}$  \\
4 arms       & 283    & $0.71$  & $-7.05$  &  $4.80\times10^{-5}$  & $4.93\times10^{-3}$  \\
5+ arms      & 286    & $0.80$  & $-8.11$  &  $7.54\times10^{-5}$  & $7.53\times10^{-3}$  \\
can't tell   & 2002   & $0.77$  & $-7.72$  &  $4.53\times10^{-5}$  & $4.54\times10^{-3}$  \\
\hline                                                                                   
tight arms   & 3239   & $0.78$  & $-7.74$  &  $5.50\times10^{-5}$  & $5.66\times10^{-3}$  \\
medium arms  & 4564   & $0.78$  & $-7.68$  &  $4.16\times10^{-5}$  & $4.21\times10^{-3}$  \\
loose arms   & 1672   & $0.78$  & $-7.64$  &  $3.21\times10^{-5}$  & $3.20\times10^{-3}$  \\
\hline                                                                                   
barred       & 3185   & $0.76$  & $-7.54$  &  $8.97\times10^{-5}$  & $8.88\times10^{-3}$  \\
unbarred     & 11746  & $0.71$  & $-6.99$  &  $3.39\times10^{-5}$  & $3.38\times10^{-3}$  \\
\hline                                                                                   
merger       & 2951   & --      & $-6.79$  &  --                                          &      --                    \\
\hline
 \end{tabular}
\caption{Basic properties of the $M_\star-SFR$ linear fit for GZ2 star-forming galaxies. $N$ is the number of galaxies with plurality classifications for spiral arm multiplicity and pitch angles, and at a cutoff of $p_{bar}=0.4$ for barred/unbarred galaxies. $\alpha$ and $\beta$ are fit according to Equation~\ref{eqn-linearfit} to data weighted by morphological vote fractions for spiral arm multiplicity and pitch angle, and to subsamples split by morphology for barred/unbarred and merging galaxies.\label{tbl-fits}}
\end{table}

The total sample analyzed in this paper consists of 48,405 star-forming galaxies. These are selected from the GZ2 spectroscopic sample with $z<0.085$ (the limit of reliable debiased morphological classification for GZ2) for galaxies classified as actively star-forming ($BPT=1$) from the MPA-JHU emission line measurements. The average color for the star-forming galaxies is relatively blue, with $(u-r)=1.6\pm0.4$. 

To parametrize the $M_\star-SFR$ relationship for the full sample of star-forming galaxies, we apply a simple linear model for the total sample and subsamples. We apply a least-squares fit where the data are weighted by the uncertainty in SFR (computed as the mean difference in the $16^{\rm th}$ and $84^{\rm th}$ percentiles from the MPA-JHU PDFs). The data are then fit to:

\begin{equation}
\log(SFR) = \alpha(\log[M_\star/M_\odot]) + \beta \hspace{20pt}[M_\odot/yr]
\label{eqn-linearfit}
\end{equation}

\noindent where $\alpha$ and $\beta$ represent the slope and offset, respectively. The formal uncertainties $\sigma_\alpha$ and $\sigma_\beta$ are taken from the covariance matrix for each least-squares fit (Table~\ref{tbl-fits}). In fitting subsamples selected by morphology, we apply the same fit to all star-forming galaxies, but weighted by the morphological likelihood in the GZ2 data. The low number of high-mass galaxies in this volume also means that we are insensitive to possible turnovers in the SFMS at $M_\star>10^{10}~M_\odot$ \citep{whi14,lee15}, emphasizing our choice to fit a linear model. 


The sample of galaxies examined here is not explicitly constructed to be volume-limited (although \citealt{bri04} estimate that SFR and $M_\star$ are essentially complete for the mass and $S/N$ limits employed). The main reason for this is that we are comparing effects between sub-samples of galaxies using the same selection functions. One possibility is that volume-limiting would deal with galaxies in which the dust content is high enough to obscure all emission lines, even in the presence of significant star formation. Such galaxies, typically [U]LIRGs, have very low space-densities at $z<0.1$, and typically lack the regular disk structure needed to categorize it for the morphologies considered here.  

In order to address volume-limiting, we have performed a detailed analysis using a series of volume-limits with upper redshift limits out to $z<0.085$ and $M_r < 20.17$. All results discussed in this paper agree with data in the volume-limited group. However, the significance of fits is smaller in the volume-limited data due to the restricted range in stellar mass ($M_\star\gtrsim10^9$), which affects the accuracy of a linear fit to the SFMS. For these reasons, we present results from the full sample of star-forming disks, which increases the sensitivity of our method to potentially small shifts between the morphologically-selected sub-samples. 


\section{Results} \label{sec-results}

We analyze the dependence of the star-forming main sequence for three different sets of disk galaxies: splitting the sample by the observed number (multiplicity) of spiral arms, the relative pitch angle (tightness or winding) of the spiral arms, and the presence of a galactic bar. Both spiral arms and galactic bars can have significant effects on the local properties of a galaxy. Dynamical effects concentrate gas in spiral arms and redistribute star formation \citep{elm86,foy10}, while longer galactic bars have redder colors and less star formation than the rest of the disk \citep{hoy11,mas12a}. We examine whether these kpc-scale effects can be seen long-term in the galaxy's SFR-mass relationship. 

\begin{figure*}
\includegraphics[angle=0,width=7.0in]{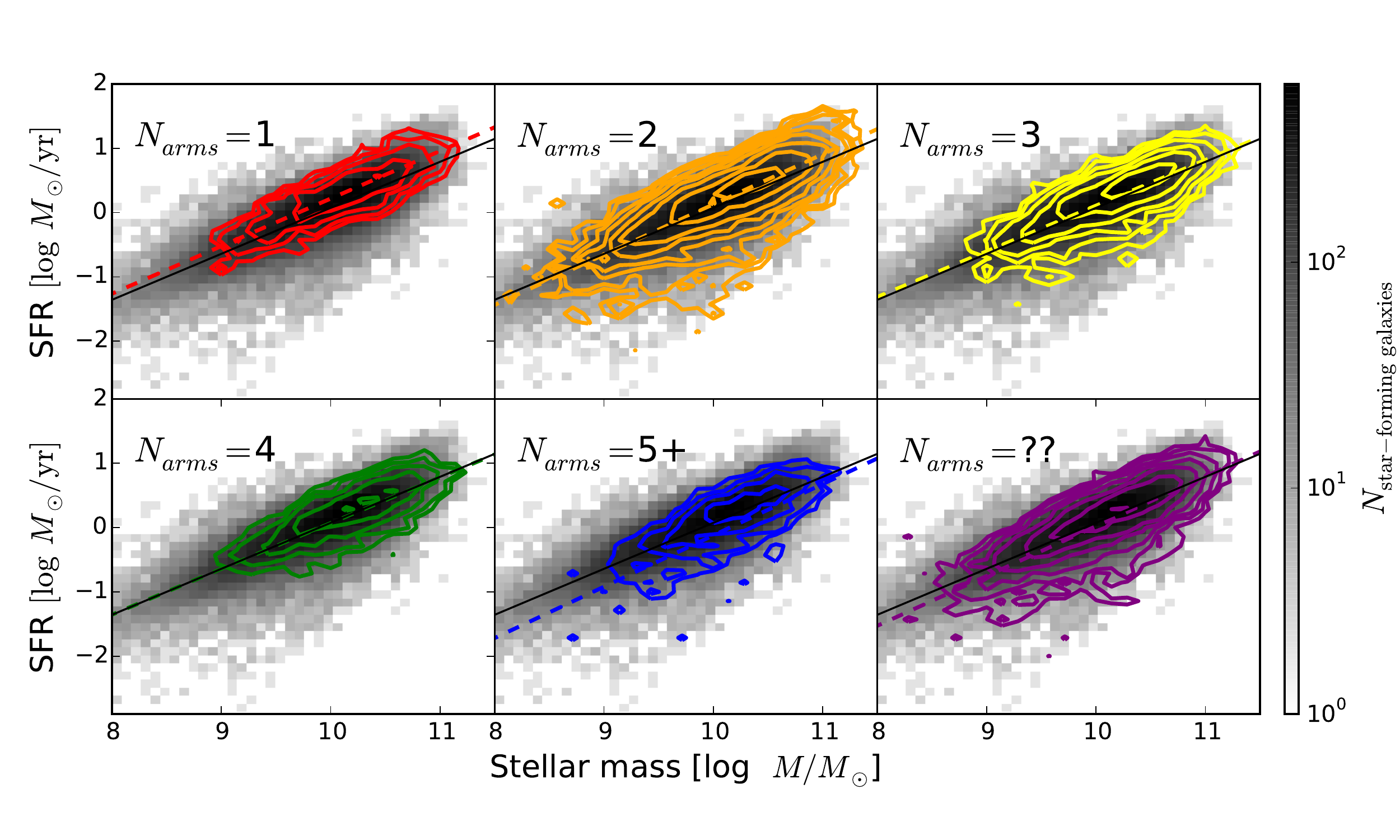}
\caption{Total star formation rate as a function of stellar mass; grayscale colors are the distribution of all star-forming galaxies in SDSS from the MPA-JHU DR7 catalog. Coloured contours in each panel show spiral galaxies weighted by the GZ2 likelihoods of hosting 1, 2, 3, 4, more than four, or ``uncertain'' numbers of spiral arms, respectively. Dotted lines show the weighted least-squares linear fit to each population as split by arm multiplicity; the solid line is the fit to all star-forming galaxies. 
\label{fig-number}}
\end{figure*}

\begin{figure*}
\includegraphics[angle=0,width=7.0in]{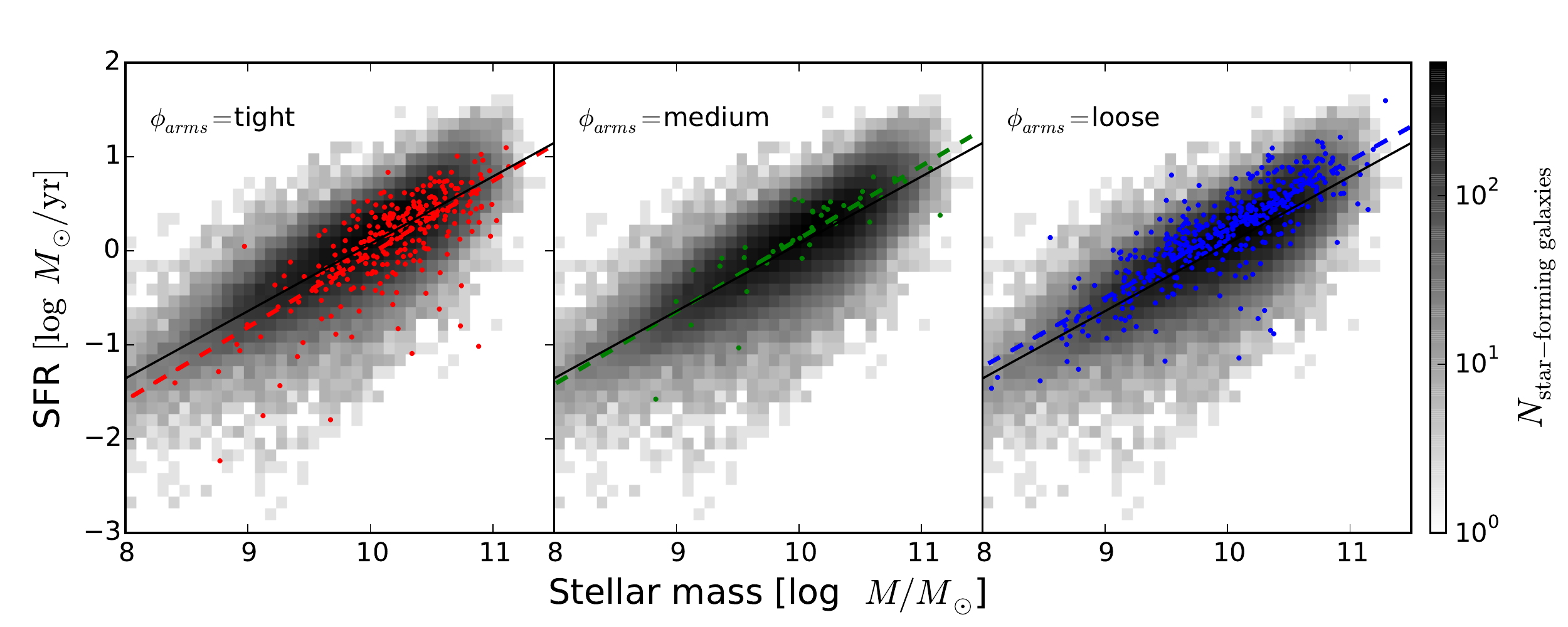}
\caption{Total star formation rate as a function of stellar mass; grayscale colors are the same as in Figure~\ref{fig-number}. From left to right: red, green, and blue points are spiral galaxies with ``tight'', ``medium'', and ``loose'' winding spiral arms as identified by GZ2 morphology flags. Dotted lines show the weighted least-squares linear fit as split by pitch angle; the solid line is the fit to all star-forming galaxies. The slight positive offset in SFR for loosely-wound spiral arms is interpreted as contamination by merging pairs of galaxies (Section~\ref{sec-discussion}).
\label{fig-winding}}
\end{figure*}

\begin{figure*}
\includegraphics[angle=0,width=7.0in]{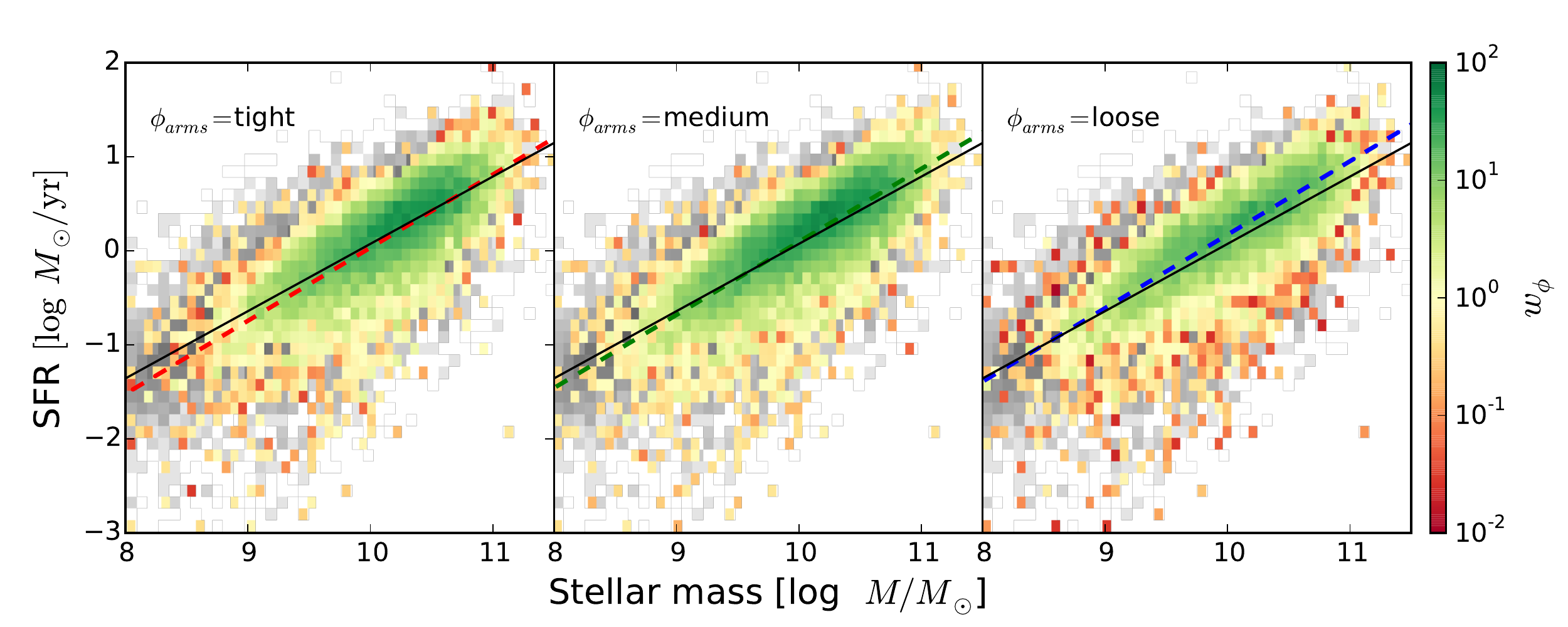}
\caption{Same as Figure~\ref{fig-winding}, but with colormaps showing all spiral galaxies weighted by the GZ2 vote fractions for ``tight'', ``medium'', and ``loose'' winding spiral arms. 
\label{fig-winding_weighted}}
\end{figure*}

\begin{figure*}
\includegraphics[angle=0,width=7.0in]{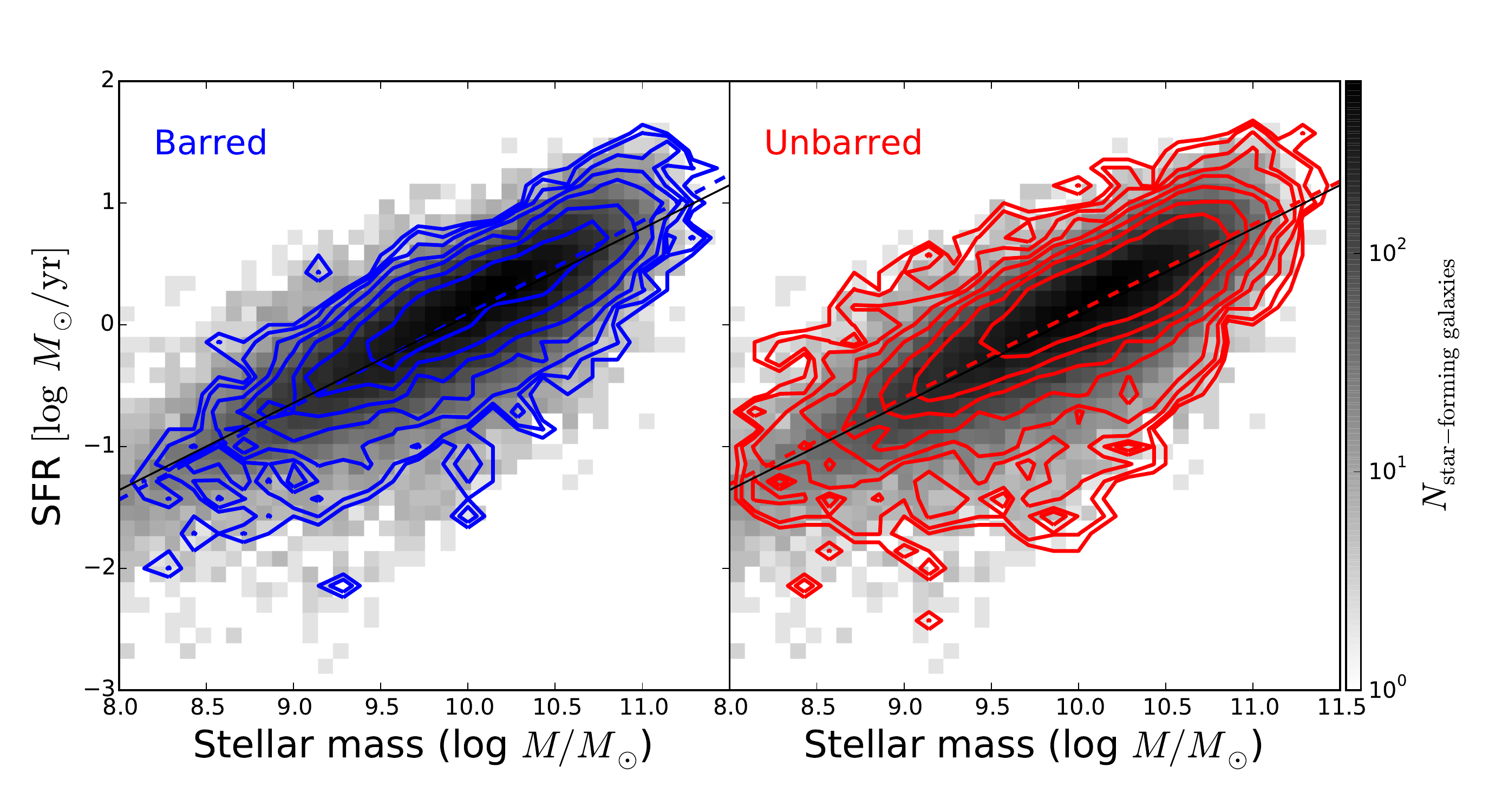}
\caption{Total star formation rate as a function of stellar mass; grayscale colors are the same as in Figure~\ref{fig-number}. Left: blue contours show the distribution of barred galaxies ($p_\textrm{bar}\ge0.4$ for previously-identified disks) from GZ2. Right: red contours are the distribution of remaining disk galaxy population with no evidence for a strong bar ($p_\textrm{bar}<0.4$). Dotted lines show the weighted least-squares linear fit to the barred/unbarred population; the solid line is the fit to all star-forming galaxies. 
\label{fig-bar}}
\end{figure*}

As a control sample, we also analyze the fits to the underlying star-forming main sequence relation for star-forming disks as measured in a sample of local SDSS galaxies (Figures~\ref{fig-number}-\ref{fig-bar}, \ref{fig-mergers}). As previously demonstrated with SDSS data \citep[e.g.,][]{bri04}, there is a tight correlation between $M_\star$ and SFR, with galaxies in the process of quenching lowering their SFR and falling below the trend. The relationship extends over at least 3~orders of magnitude in both $M_\star$ and SFR. Fits to the SFMS for all subsamples in this paper are listed in Table~\ref{tbl-fits}.

Figure~\ref{fig-number} shows the SFR as a function of $M_\star$ for disk galaxies separated by their arm multiplicity. The GZ2 data separates disk galaxies with visible spiral arms into categories of 1, 2, 3, 4, or more than four spiral arms; there is also an additional option if the number of spiral arms cannot be accurately determined. For this analysis, galaxies in each mass/SFR bin is weighted by the vote fraction for the morphology being tested. The fit to the star-forming main sequence for all six sub-samples of spiral galaxies split by arm multiplicity tightly follows that of the total star-forming disk population. Both the slopes and offsets of each linear fit are consistent within the formal fitting errors in Table~\ref{tbl-fits}.

One-armed spiral galaxies present an interesting case, with the fit to the weighted population lying slightly \emph{above} that of all star-forming spirals. This is consistent with the work of \citet{cas13}, who showed that one-armed spirals in GZ2 are robust indicators of close interactions at projected distances of $r_p < 50~h^{-1}$~kpc. The underlying reason is that many ``one-armed spirals'' are in fact caused by bridges or tidal tails from interactions with a nearby companion instead of secular processes. We discuss the likely role of merging/interacting galaxies in Section~\ref{sec-discussion} (also see Figure~\ref{fig-mergers}).

The only morphologies that extend slightly below the SFMS are those with the highest level of multiplicity (five or more arms). The best-fit line for this population has a steeper slope, driven by the galaxies with relatively low SFR at $10^9<M/M_\star<10^{10}$. This is a new and unusual result; one possible explanation is that stellar disks in the process of quenching will have fewer bright \mbox{H\,{\sc ii}}~regions and the contrast between the arm and interarm regions is increased. This could result in better visibility for older (and potentially overlapping) spatial modes in the galaxy's disk, increasing the measured multiplicity. It should be emphasized, though, that 5+-arm spirals represent the smallest morphological group in the sample, and that the associated fit errors in Table~\ref{tbl-fits} are the largest for any multiplicity; simple statistical variance cannot be ruled out as an explanation for the best-fit line to the SFMS.

We have repeated the analysis above for the sub-sample of disk galaxies for which the spiral multiplicity is determined with high confidence ($p_\textrm{arms~number} > 0.8$) by GZ2, thus eliminating ``intermediate'' galaxies for which the morphology is uncertain. These galaxies ($N=10,035$) are dominated by two-armed spirals, which are the only spiral multiplicity for which significant numbers of galaxies at $M_\star<10^9~M_\odot$ are detected. The results for the SFMS are qualitatively the same as when using the weighting scheme, although we note that there are too few examples ($N<10$) of either three- or four-armed spirals for a reliable fit. The offset of the one-armed spirals above the SFMS is also significantly more pronounced when using high-confidence morphologies.

The pitch angle of the spiral arms also has no significant change on the star-forming main sequence relation (Figure~\ref{fig-winding}). We separate galaxies by their relative pitch angles (defined as ``tight'', ``medium'', and ``loose''); the pitch angle is typically used as one of the primary parameters for separating galaxies along the Hubble tuning fork. \citet{wil13} show, however, that pitch angle only weakly correlates with Hubble type from expert visual classifications, and that the bulge-to-disk ratio is a more important driver. There is no significant shift with respect to the SFMS as a function of pitch angle for spiral galaxies. The small increase above the main sequence for loosely-wound galaxies is also consistent with \citet{cas13}, who show that this morphology also correlates with close pairs and interactions. 

It should also be noted that the galaxies in GZ2 flagged as a function of pitch angle are not representative of the true vote distribution. The points in Figure~\ref{fig-winding} would suggest that there are relatively few spiral galaxies overall, and that most are either tightly or loosely wound. In fact, the plurality classification for most galaxies is for medium-winding; the spread in votes is typically large, though, and so users rarely agree on the ``medium'' option at the 80\% level which sets the flag. An alternative method is to analyze the morphology of spiral galaxies by directly weighting them as a function of the pitch angle categories (Figure~\ref{fig-winding_weighted}), which has the advantage of including all spiral galaxies. This shows an even tighter agreement between the samples separated by pitch angle and that of the full star-forming sample.

Finally, we examine the effect of a large-scale galactic bar on the star-forming main sequence. This sample has significantly more galaxies than those including spiral arm morphology, since the classification is at a higher level in the GZ2 tree and has only two choices. This results in a higher percentage of consensus classifications in the GZ2 catalog. Figure~\ref{fig-bar} shows the SFMS for both barred and unbarred galaxies. Although the fraction of barred galaxies varies as a function of stellar mass \citep{she08a,cam10,mas11c,che13}, both the linear fits and ranges of the sub-populations are consistent with all star-forming galaxies. In other words, the presence of a bar does not affect a star-forming galaxy's position on the SFMS. 

\begin{figure}
\includegraphics[angle=0,width=3.5in]{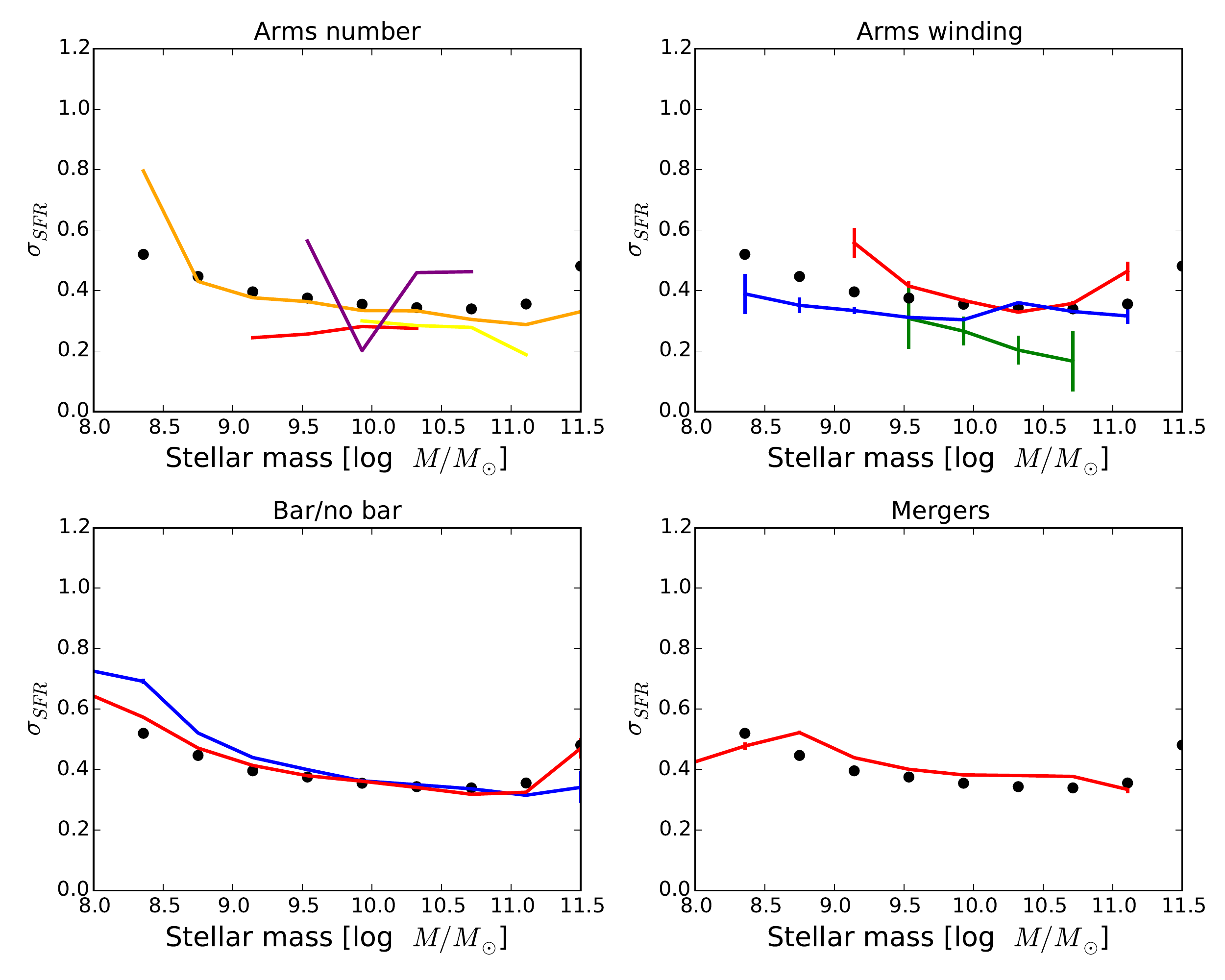}
\caption{Width of the star-forming main sequence ($\sigma_{SFR}$) as a function of stellar mass, as measured by the sample standard deviation. Black points represent the entire star-forming population. Disk subsamples are overplotted as solid lines; colours are the same as the respective plots in Figures~\ref{fig-number}, \ref{fig-winding}, \ref{fig-bar} and \ref{fig-mergers}. Morphological categories or mass ranges with fewer than 10 galaxies/bin are not plotted; this includes all galaxies with 3, 4, and 5+ spiral arms.
\label{fig-sigma}}
\end{figure}

The agreement of all sub-varieties of star-forming galaxies is supported by the close agreement to the linear fits to the data for all well-sampled categories (Table~\ref{tbl-fits}). This tracks only the slope and offset of the distribution, however, and not its width. We thus also compare the sample standard deviation ($\sigma_{SFR}$) to the star-forming galaxy population over its various morphological subsamples. The value of $\sigma_{SFR}$ monotonically decreases with increasing mass over the range $8.0<\log(M/M_\odot)<11.5$. For all morphological populations examined in this paper, the widths of their distributions are consistent with the broader population (Figure~\ref{fig-sigma}). 

We have also examined all the populations of galaxies described above (bars, arm pitch angle, arm multiplicity) and measured the differences when using specific star formation rate ($\textrm{sSFR}\equiv\textrm{SFR}/M_\star$) instead of SFR. There is no significant change in any of the morphologically-selected categories as compared to the general SFMS. 


\section{Discussion}\label{sec-discussion}

\begin{figure*}
\includegraphics[angle=0,width=7.0in]{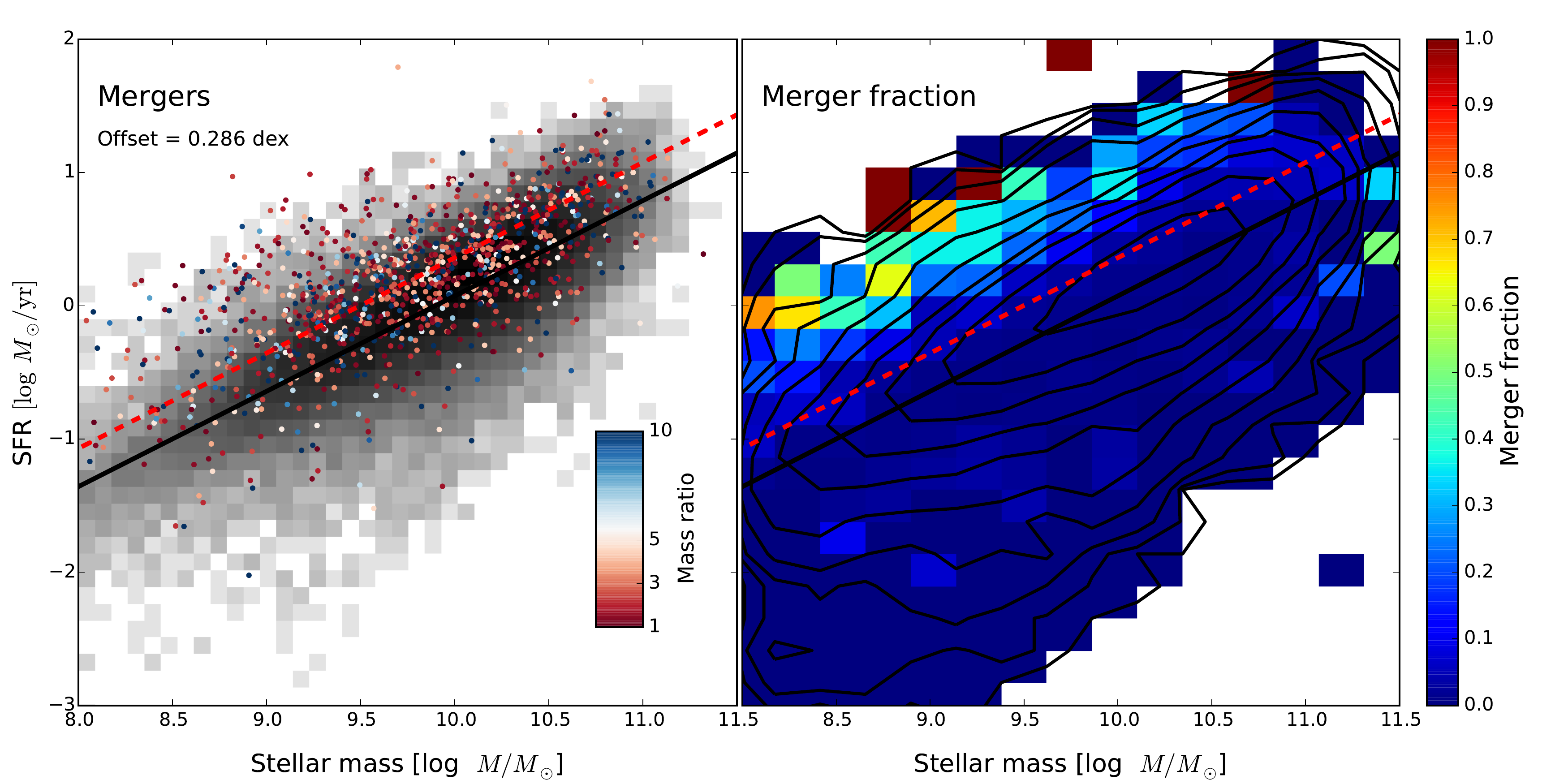}
\caption{Total star formation rate as a function of stellar mass; grayscale colors are the same as in Figure~\ref{fig-number}. Left: Coloured points show 2,978~merging galaxies from \citet{dar10a}. Mergers are colour-coded by the mass ratio of the primary and secondary galaxies; there is no clear difference in the merging populations with regard to the SFMS when comparing major to minor mergers. When fixing the slope of the star-forming main sequence and allowing the offset to vary, mergers (dotted line) have higher SFRs by $\sim0.3$~dex compared to all star-forming galaxies (solid line). Right: Star-forming galaxies binned and colour-coded by merger fraction ($N_\textrm{mergers}/N_\textrm{star-forming galaxies})$. Overplotted lines are the same as left plot. Of the galaxies that lie furthest above the SFMS, more than $50\%$ are unambiguous mergers. 
\label{fig-mergers}}
\end{figure*}

Our results show that the star-forming main sequence is remarkably robust to the details of the spatial distribution of star formation \textit{within} galaxies. Testing for a wide range of morphological sub-types of star-forming disk galaxies yields no statistically significant difference in the relative position of these sub-types vis-\'a-vis the main sequence. Neither the number or pitch angle of spiral arms, or the presence of a large-scale bar are correlated with any detectable increase or decrease in the efficiency of star formation. The system which regulates star formation in galaxies is thus either not affected by the details of the spatial distribution of star formation, or its regulatory effect is so strong that it wipes out any such effect in a short time. 

\citet{abr14} found that by normalizing galaxies by the stellar mass of the disk alone, the slope of the SFMS is consistent with only a linear trend (removing any dependence on mass). Although this correction to the disk stellar mass homogenizes the SFMS for disks with a range of $B/T$, the intrinsic dispersion ($\sigma_{SFR}$) of the sequence must be a result of contributions by bars, disk dynamics, halo heating, AGN activity, environment and/or gas accretion history, among other factors \citep{dut10}. Our results show that the neither of the first two factors play dominant roles in controlling $\sigma_{SFR}$, at least as far as major dynamical drivers (such as strong bars or additional arms) are concerned. Thus while the overall bulge strength does affect the position of a galaxy on the SFMS \citep{mar09,che12a,fan13,kav14,lan14,oma14}, the structure of the \emph{disk itself} does not. This is also consistent with recent models in which details of the feedback, which also relate strongly to the galaxy properties, have little effect on the SFMS \citep{hop14}. Alternatively, this also agrees with models in which the SFMS is the result of stochastic processes, rather than deterministic physics related to galaxy evolution \citep{kel14b}. 

The lack of any difference in SFR as a function of mass for barred vs. unbarred galaxies is in general agreement with \citet{ell11}, who find an increase of $\Delta\textrm{SFR}\sim0.15$~dex, but only for galaxies with $M_\star>10^{10.7}~M_\odot$. This is at the very upper end of the mass range probed in our analysis of barred vs. unbarred star-forming disks (Figure~\ref{fig-bar}). If the increase in star formation is limited to the central kiloparsec of the disk (as demonstrated using fibre SFR measurements), an increase in possible bar-driven SFR increase is seen down to $M_\star=10^{10}~M_\odot$. 

The absence of an apparent influence of the bar on the SFMS is still at apparent odds with the anti-correlation between atomic gas mass fraction and the presence of a bar \citep{mas12a}. One possible explanation is that strong bars are driven by spiral modes with star formation proceeding radially outward from the center; in that case, the influence of the bar may not be seen in star-formation diagnostics averaged over the entire disk of the galaxy (as used in this paper). It is also important to note that the selection of only \emph{star-forming} disk galaxies for this study excludes passive disks, which are known to be significantly redder and more massive than their star-forming counterparts \citep{mas10a,cor12}. 

Amongst individual galaxies that lie significantly off the SFMS, compact starburst galaxies show the largest increase in SFR at a given mass \citep{elb11}. In the local Universe, these include optically-identified ``green pea'' galaxies, which have unusually high sSFR and can lie more than 1~dex above the SFMS \citep{car09}. While few green pea galaxies have detailed imaging available, their most common morphology is in a clumpy arrangement with knots of bright star formation. There is thus little evidence for a dynamically-settled disk (in any arrangment) for galaxies in the local Universe lying significantly above the SFMS. 

As a comparison to the kpc-scale structures discussed above, we analyze the impact of the most significant forcing event to a galaxy system known -- a major galaxy merger (Figure~\ref{fig-mergers}). In these systems, which are in various stages of coalescence, star formation rates are increased by only an average of 0.29~dex (less than a factor of two). \citet{dar10} showed that at $z<0.1$, galaxies with intense bursts of star formation are limited to only the spiral (disc) galaxies. This increase in star formation for mergers does show a strong evolution in redshift out to at least $1.5<z<2.5$, likely due to the higher gas fractions involved \citep{dad10,rod11}. The projected separation between galaxies in which this occurs, based on both the observed merger fraction and sSFR is $\sim0.1$~Mpc$/h$ \citep{ski09}. Our measurements are consistent with observations of galaxies at $z\simeq2$ \citep{kav13b}, which support a merger-driven increase of only a factor of $\sim2$ in sSFR. 

The location of mergers on the present-day SFMS shows just how stable the regulatory system in galaxies really is. Almost all simulations of galaxy mergers predict a steep increase in the star formation rate during both first passage and final coalescence \citep[e.g.,][]{hop08}. The magnitude of this increase often depends on the details of the simulation, but can range from factors of 10 to 100. Observations of mergers in Stripe~82 data, however, limit this increase to between factors of 2 and 6 \citep{kav14}. In the low-redshift universe sampled by SDSS and Galaxy~Zoo, we find no evidence for an enhancement more than an order of magnitude. This in turn suggests that the current generation of galaxy merger simulations misses critical feedback mechanisms that prevent runaway peaks in star formation rates during mergers.

\section{Conclusions}

We analyze for the first time the detailed structure of disks in large samples of star-forming galaxies in the local Universe as related to their position on the $M_\star$-SFR relation. This analysis is made possible by using morphological classifications from the Galaxy~Zoo~2 project. We find that neither the slope nor the dispersion of the star-forming galaxies are affected when splitting the sample into different categories of disks, including barred/unbarred galaxies, the pitch angle of spiral arms, or the number of spiral arms. 

The uniformity of disk galaxies along the SFMS, regardless of their kpc-scale structure, argues for the system as a whole being strongly self-regulated. While smaller regions of the galaxy can experience (likely temporary) increases in star formation, the amount of star formation in the disk as a whole is conserved. This is preserved even for the strongest forcing events, including major mergers; the physics governing the SFMS are primarily driven by the overall mass of the system. This means that simulations of galaxy evolution must be able to meet the challenge of reproducing the wide range of disk morphologies observed along the Hubble sequence (and in various merger stages) while simultaneously managing feedback so that \emph{all} disk types maintain the same tight relationship to the SFMS.


\section*{Acknowledgments}

The data in this paper are the result of the efforts of the Galaxy~Zoo volunteers, without whom none of this work would be possible. Their efforts are individually acknowledged at \url{http://authors.galaxyzoo.org}. 

We thank Rory Smith, Lucio Mayer, and Bruce Elmegreen for useful discussions. This research made use of TOPCAT, an interactive graphical viewer and editor for tabular data \citep{tay05} and Astropy, a community-developed core Python package for astronomy \citep{ast13}. The development of Galaxy~Zoo~2 was supported by The Leverhulme Trust. KWW and LFF are supported by the US National Science Foundation under grant DRL-0941610. KS gratefully acknowledges support from Swiss National Science Foundation Grant PP00P2\_138979/1.

Funding for the SDSS and SDSS-II has been provided by the Alfred P. Sloan Foundation, the Participating Institutions, the National Science Foundation, the U.S. Department of Energy, the National Aeronautics and Space Administration, the Japanese Monbukagakusho, the Max Planck Society, and the Higher Education Funding Council for England. The SDSS website is \url{http://www.sdss.org/}.

The SDSS is managed by the Astrophysical Research Consortium for the Participating Institutions. The Participating Institutions are the American Museum of Natural History, Astrophysical Institute Potsdam, University of Basel, University of Cambridge, Case Western Reserve University, University of Chicago, Drexel University, Fermilab, the Institute for Advanced Study, the Japan Participation Group, Johns Hopkins University, the Joint Institute for Nuclear Astrophysics, the Kavli Institute for Particle Astrophysics and Cosmology, the Korean Scientist Group, the Chinese Academy of Sciences (LAMOST), Los Alamos National Laboratory, the Max-Planck-Institute for Astronomy (MPIA), the Max-Planck-Institute for Astrophysics (MPA), New Mexico State University, Ohio State University, University of Pittsburgh, University of Portsmouth, Princeton University, the United States Naval Observatory, and the University of Washington.


\bibliography{kwrefs}

\begin{thebibliography}{57}
\expandafter\ifx\csname natexlab\endcsname\relax\def\natexlab#1{#1}\fi

\bibitem[{{Abazajian} {et~al}\mbox{.}(2009){Abazajian}, {Adelman-McCarthy},
  {Ag{\"u}eros}, {Allam}, {Allende Prieto}, {An}, {Anderson}, {Anderson},
  {Annis}, {Bahcall}, \& et~al.}]{aba09}
{Abazajian} K.~N. {et~al.}, 2009, \apjs, 182, 543

\bibitem[{{Abramson} {et~al}\mbox{.}(2014){Abramson}, {Kelson}, {Dressler},
  {Poggianti}, {Gladders}, {Oemler}, \& {Vulcani}}]{abr14}
{Abramson} L.~E., {Kelson} D.~D., {Dressler} A., {Poggianti} B., {Gladders}
  M.~D., {Oemler}, Jr. A., {Vulcani} B., 2014, \apjl, 785, L36

\bibitem[{{Astropy Collaboration} {et~al}\mbox{.}(2013){Astropy Collaboration},
  {Robitaille}, {Tollerud}, {Greenfield}, {Droettboom}, {Bray}, {Aldcroft},
  {Davis}, {Ginsburg}, {Price-Whelan}, {Kerzendorf}, {Conley}, {Crighton},
  {Barbary}, {Muna}, {Ferguson}, {Grollier}, {Parikh}, {Nair}, {Unther},
  {Deil}, {Woillez}, {Conseil}, {Kramer}, {Turner}, {Singer}, {Fox}, {Weaver},
  {Zabalza}, {Edwards}, {Azalee Bostroem}, {Burke}, {Casey}, {Crawford},
  {Dencheva}, {Ely}, {Jenness}, {Labrie}, {Lim}, {Pierfederici}, {Pontzen},
  {Ptak}, {Refsdal}, {Servillat}, \& {Streicher}}]{ast13}
{Astropy Collaboration} {et~al.}, 2013, \aap, 558, A33

\bibitem[{{Baldwin} {et~al}\mbox{.}(1981){Baldwin}, {Phillips}, \&
  {Terlevich}}]{bal81}
{Baldwin} J.~A., {Phillips} M.~M., {Terlevich} R., 1981, \pasp, 93, 5

\bibitem[{{Bamford} {et~al}\mbox{.}(2009){Bamford}, {Nichol}, {Baldry}, {Land},
  {Lintott}, {Schawinski}, {Slosar}, {Szalay}, {Thomas}, {Torki}, {Andreescu},
  {Edmondson}, {Miller}, {Murray}, {Raddick}, \& {Vandenberg}}]{bam09}
{Bamford} S.~P. {et~al.}, 2009, \mnras, 393, 1324

\bibitem[{{Bouch{\'e}} {et~al}\mbox{.}(2010){Bouch{\'e}}, {Dekel}, {Genzel},
  {Genel}, {Cresci}, {F{\"o}rster Schreiber}, {Shapiro}, {Davies}, \&
  {Tacconi}}]{bou10}
{Bouch{\'e}} N. {et~al.}, 2010, \apj, 718, 1001

\bibitem[{{Brinchmann} {et~al}\mbox{.}(2004){Brinchmann}, {Charlot}, {White},
  {Tremonti}, {Kauffmann}, {Heckman}, \& {Brinkmann}}]{bri04}
{Brinchmann} J., {Charlot} S., {White} S.~D.~M., {Tremonti} C., {Kauffmann} G.,
  {Heckman} T., {Brinkmann} J., 2004, \mnras, 351, 1151

\bibitem[{{Cameron} {et~al}\mbox{.}(2010){Cameron}, {Carollo}, {Oesch},
  {Aller}, {Bschorr}, {Cerulo}, {Aussel}, {Capak}, {Le Floc'h}, {Ilbert},
  {Kneib}, {Koekemoer}, {Leauthaud}, {Lilly}, {Massey}, {McCracken}, {Rhodes},
  {Salvato}, {Sanders}, {Scoville}, {Sheth}, {Taniguchi}, \&
  {Thompson}}]{cam10}
{Cameron} E. {et~al.}, 2010, \mnras, 409, 346

\bibitem[{{Cardamone} {et~al}\mbox{.}(2009){Cardamone}, {Schawinski}, {Sarzi},
  {Bamford}, {Bennert}, {Urry}, {Lintott}, {Keel}, {Parejko}, {Nichol},
  {Thomas}, {Andreescu}, {Murray}, {Raddick}, {Slosar}, {Szalay}, \&
  {Vandenberg}}]{car09}
{Cardamone} C. {et~al.}, 2009, \mnras, 399, 1191

\bibitem[{{Casteels} {et~al}\mbox{.}(2013){Casteels}, {Bamford}, {Skibba},
  {Masters}, {Lintott}, {Keel}, {Schawinski}, {Nichol}, \& {Smith}}]{cas13}
{Casteels} K.~R.~V. {et~al.}, 2013, \mnras, 429, 1051

\bibitem[{{Cheung} {et~al}\mbox{.}(2013){Cheung}, {Athanassoula}, {Masters},
  {Nichol}, {Bosma}, {Bell}, {Faber}, {Koo}, {Lintott}, {Melvin}, {Schawinski},
  {Skibba}, \& {Willett}}]{che13}
{Cheung} E. {et~al.}, 2013, \apj, 779, 162

\bibitem[{{Cheung} {et~al}\mbox{.}(2012){Cheung}, {Faber}, {Koo}, {Dutton},
  {Simard}, {McGrath}, {Huang}, {Bell}, {Dekel}, {Fang}, {Salim}, {Barro},
  {Bundy}, {Coil}, {Cooper}, {Conselice}, {Davis}, {Dom{\'{\i}}nguez},
  {Kassin}, {Kocevski}, {Koekemoer}, {Lin}, {Lotz}, {Newman}, {Phillips},
  {Rosario}, {Weiner}, \& {Willmer}}]{che12a}
{Cheung} E. {et~al.}, 2012, \apj, 760, 131

\bibitem[{{Cortese}(2012)}]{cor12}
{Cortese} L., 2012, \aap, 543, A132

\bibitem[{{Daddi} {et~al}\mbox{.}(2007){Daddi}, {Dickinson}, {Morrison},
  {Chary}, {Cimatti}, {Elbaz}, {Frayer}, {Renzini}, {Pope}, {Alexander},
  {Bauer}, {Giavalisco}, {Huynh}, {Kurk}, \& {Mignoli}}]{dad07}
{Daddi} E. {et~al.}, 2007, \apj, 670, 156

\bibitem[{{Daddi} {et~al}\mbox{.}(2010){Daddi}, {Elbaz}, {Walter}, {Bournaud},
  {Salmi}, {Carilli}, {Dannerbauer}, {Dickinson}, {Monaco}, \&
  {Riechers}}]{dad10}
{Daddi} E. {et~al.}, 2010, \apjl, 714, L118

\bibitem[{{Darg} {et~al}\mbox{.}(2010{\natexlab{a}}){Darg}, {Kaviraj},
  {Lintott}, {Schawinski}, {Sarzi}, {Bamford}, {Silk}, {Andreescu}, {Murray},
  {Nichol}, {Raddick}, {Slosar}, {Szalay}, {Thomas}, \& {Vandenberg}}]{dar10}
{Darg} D.~W. {et~al.}, 2010{\natexlab{a}}, \mnras, 401, 1552

\bibitem[{{Darg} {et~al}\mbox{.}(2010{\natexlab{b}}){Darg}, {Kaviraj},
  {Lintott}, {Schawinski}, {Sarzi}, {Bamford}, {Silk}, {Proctor}, {Andreescu},
  {Murray}, {Nichol}, {Raddick}, {Slosar}, {Szalay}, {Thomas}, \&
  {Vandenberg}}]{dar10a}
{Darg} D.~W. {et~al.}, 2010{\natexlab{b}}, \mnras, 401, 1043

\bibitem[{{De~Lucia} {et~al}\mbox{.}(2012){De~Lucia}, {Weinmann}, {Poggianti},
  {Arag{\'o}n-Salamanca}, \& {Zaritsky}}]{del12}
{De~Lucia} G., {Weinmann} S., {Poggianti} B.~M., {Arag{\'o}n-Salamanca} A.,
  {Zaritsky} D., 2012, \mnras, 423, 1277

\bibitem[{{Dutton} {et~al}\mbox{.}(2010){Dutton}, {van den Bosch}, \&
  {Dekel}}]{dut10}
{Dutton} A.~A., {van den Bosch} F.~C., {Dekel} A., 2010, \mnras, 405, 1690

\bibitem[{{Elbaz} {et~al}\mbox{.}(2011){Elbaz}, {Dickinson}, {Hwang},
  {D{\'{\i}}az-Santos}, {Magdis}, {Magnelli}, {Le Borgne}, {Galliano},
  {Pannella}, {Chanial}, {Armus}, {Charmandaris}, {Daddi}, {Aussel}, {Popesso},
  {Kartaltepe}, {Altieri}, {Valtchanov}, {Coia}, {Dannerbauer}, {Dasyra},
  {Leiton}, {Mazzarella}, {Alexander}, {Buat}, {Burgarella}, {Chary}, {Gilli},
  {Ivison}, {Juneau}, {Le Floc'h}, {Lutz}, {Morrison}, {Mullaney}, {Murphy},
  {Pope}, {Scott}, {Brodwin}, {Calzetti}, {Cesarsky}, {Charlot}, {Dole},
  {Eisenhardt}, {Ferguson}, {F{\"o}rster Schreiber}, {Frayer}, {Giavalisco},
  {Huynh}, {Koekemoer}, {Papovich}, {Reddy}, {Surace}, {Teplitz}, {Yun}, \&
  {Wilson}}]{elb11}
{Elbaz} D. {et~al.}, 2011, \aap, 533, A119

\bibitem[{{Ellison} {et~al}\mbox{.}(2011){Ellison}, {Nair}, {Patton},
  {Scudder}, {Mendel}, \& {Simard}}]{ell11}
{Ellison} S.~L., {Nair} P., {Patton} D.~R., {Scudder} J.~M., {Mendel} J.~T.,
  {Simard} L., 2011, \mnras, 416, 2182

\bibitem[{{Elmegreen} \& {Elmegreen}(1986)}]{elm86}
{Elmegreen} B.~G., {Elmegreen} D.~M., 1986, \apj, 311, 554

\bibitem[{{Fang} {et~al}\mbox{.}(2013){Fang}, {Faber}, {Koo}, \&
  {Dekel}}]{fan13}
{Fang} J.~J., {Faber} S.~M., {Koo} D.~C., {Dekel} A., 2013, \apj, 776, 63

\bibitem[{{Foyle} {et~al}\mbox{.}(2010){Foyle}, {Rix}, {Walter}, \&
  {Leroy}}]{foy10}
{Foyle} K., {Rix} H.-W., {Walter} F., {Leroy} A.~K., 2010, \apj, 725, 534

\bibitem[{{Hinshaw} {et~al}\mbox{.}(2013){Hinshaw}, {Larson}, {Komatsu},
  {Spergel}, {Bennett}, {Dunkley}, {Nolta}, {Halpern}, {Hill}, {Odegard},
  {Page}, {Smith}, {Weiland}, {Gold}, {Jarosik}, {Kogut}, {Limon}, {Meyer},
  {Tucker}, {Wollack}, \& {Wright}}]{hin13}
{Hinshaw} G. {et~al.}, 2013, \apjs, 208, 19

\bibitem[{{Hopkins} {et~al}\mbox{.}(2008){Hopkins}, {Hernquist}, {Cox}, \&
  {Kere{\v s}}}]{hop08}
{Hopkins} P.~F., {Hernquist} L., {Cox} T.~J., {Kere{\v s}} D., 2008, \apjs,
  175, 356

\bibitem[{{Hopkins} {et~al}\mbox{.}(2014){Hopkins}, {Kere{\v s}}, {O{\~n}orbe},
  {Faucher-Gigu{\`e}re}, {Quataert}, {Murray}, \& {Bullock}}]{hop14}
{Hopkins} P.~F., {Kere{\v s}} D., {O{\~n}orbe} J., {Faucher-Gigu{\`e}re} C.-A.,
  {Quataert} E., {Murray} N., {Bullock} J.~S., 2014, \mnras, 445, 581

\bibitem[{{Hoyle} {et~al}\mbox{.}(2011){Hoyle}, {Masters}, {Nichol},
  {Edmondson}, {Smith}, {Lintott}, {Scranton}, {Bamford}, {Schawinski}, \&
  {Thomas}}]{hoy11}
{Hoyle} B. {et~al.}, 2011, \mnras, 415, 3627

\bibitem[{{Kauffmann} {et~al}\mbox{.}(2003{\natexlab{a}}){Kauffmann},
  {Heckman}, {Tremonti}, {Brinchmann}, {Charlot}, {White}, {Ridgway},
  {Brinkmann}, {Fukugita}, {Hall}, {Ivezi{\'c}}, {Richards}, \&
  {Schneider}}]{kau03}
{Kauffmann} G. {et~al.}, 2003{\natexlab{a}}, \mnras, 346, 1055

\bibitem[{{Kauffmann} {et~al}\mbox{.}(2003{\natexlab{b}}){Kauffmann},
  {Heckman}, {White}, {Charlot}, {Tremonti}, {Brinchmann}, {Bruzual}, {Peng},
  {Seibert}, {Bernardi}, {Blanton}, {Brinkmann}, {Castander}, {Cs{\'a}bai},
  {Fukugita}, {Ivezic}, {Munn}, {Nichol}, {Padmanabhan}, {Thakar}, {Weinberg},
  \& {York}}]{kau03a}
{Kauffmann} G. {et~al.}, 2003{\natexlab{b}}, \mnras, 341, 33

\bibitem[{{Kaviraj}(2014)}]{kav14}
{Kaviraj} S., 2014, \mnras, 440, 2944

\bibitem[{{Kaviraj} {et~al}\mbox{.}(2013){Kaviraj}, {Cohen}, {Windhorst},
  {Silk}, {O'Connell}, {Dopita}, {Dekel}, {Hathi}, {Straughn}, \&
  {Rutkowski}}]{kav13b}
{Kaviraj} S. {et~al.}, 2013, \mnras, 429, L40

\bibitem[{{Kelson}(2014)}]{kel14b}
{Kelson} D.~D., 2014, ArXiv e-prints, 1406.5191

\bibitem[{{Lang} {et~al}\mbox{.}(2014){Lang}, {Wuyts}, {Somerville},
  {F{\"o}rster Schreiber}, {Genzel}, {Bell}, {Brammer}, {Dekel}, {Faber},
  {Ferguson}, {Grogin}, {Kocevski}, {Koekemoer}, {Lutz}, {McGrath}, {Momcheva},
  {Nelson}, {Primack}, {Rosario}, {Skelton}, {Tacconi}, {van Dokkum}, \&
  {Whitaker}}]{lan14}
{Lang} P. {et~al.}, 2014, \apj, 788, 11

\bibitem[{{Lee} {et~al}\mbox{.}(2015){Lee}, {Sanders}, {Casey}, {Toft},
  {Scoville}, {Hung}, {Le Floc'h}, {Ilbert}, {Zahid}, {Aussel}, {Capak},
  {Kartaltepe}, {Kewley}, {Li}, {Schawinski}, {Sheth}, \& {Xiao}}]{lee15}
{Lee} N. {et~al.}, 2015, ArXiv e-prints, 1501.01080

\bibitem[{{Lilly} {et~al}\mbox{.}(2013){Lilly}, {Carollo}, {Pipino}, {Renzini},
  \& {Peng}}]{lil13}
{Lilly} S.~J., {Carollo} C.~M., {Pipino} A., {Renzini} A., {Peng} Y., 2013,
  \apj, 772, 119

\bibitem[{{Lintott} {et~al}\mbox{.}(2011){Lintott}, {Schawinski}, {Bamford},
  {Slosar}, {Land}, {Thomas}, {Edmondson}, {Masters}, {Nichol}, {Raddick},
  {Szalay}, {Andreescu}, {Murray}, \& {Vandenberg}}]{lin11}
{Lintott} C. {et~al.}, 2011, \mnras, 410, 166

\bibitem[{{Lintott} {et~al}\mbox{.}(2008){Lintott}, {Schawinski}, {Slosar},
  {Land}, {Bamford}, {Thomas}, {Raddick}, {Nichol}, {Szalay}, {Andreescu},
  {Murray}, \& {Vandenberg}}]{lin08}
{Lintott} C.~J. {et~al.}, 2008, \mnras, 389, 1179

\bibitem[{{Martig} {et~al}\mbox{.}(2009){Martig}, {Bournaud}, {Teyssier}, \&
  {Dekel}}]{mar09}
{Martig} M., {Bournaud} F., {Teyssier} R., {Dekel} A., 2009, \apj, 707, 250

\bibitem[{{Masters} {et~al}\mbox{.}(2010){Masters}, {Mosleh}, {Romer},
  {Nichol}, {Bamford}, {Schawinski}, {Lintott}, {Andreescu}, {Campbell},
  {Crowcroft}, {Doyle}, {Edmondson}, {Murray}, {Raddick}, {Slosar}, {Szalay},
  \& {Vandenberg}}]{mas10a}
{Masters} K.~L. {et~al.}, 2010, \mnras, 405, 783

\bibitem[{{Masters} {et~al}\mbox{.}(2012){Masters}, {Nichol}, {Haynes}, {Keel},
  {Lintott}, {Simmons}, {Skibba}, {Bamford}, {Giovanelli}, \&
  {Schawinski}}]{mas12a}
{Masters} K.~L. {et~al.}, 2012, \mnras, 424, 2180

\bibitem[{{Masters} {et~al}\mbox{.}(2011){Masters}, {Nichol}, {Hoyle},
  {Lintott}, {Bamford}, {Edmondson}, {Fortson}, {Keel}, {Schawinski}, {Smith},
  \& {Thomas}}]{mas11c}
{Masters} K.~L. {et~al.}, 2011, \mnras, 411, 2026

\bibitem[{{Noeske} {et~al}\mbox{.}(2007){Noeske}, {Weiner}, {Faber},
  {Papovich}, {Koo}, {Somerville}, {Bundy}, {Conselice}, {Newman},
  {Schiminovich}, {Le Floc'h}, {Coil}, {Rieke}, {Lotz}, {Primack}, {Barmby},
  {Cooper}, {Davis}, {Ellis}, {Fazio}, {Guhathakurta}, {Huang}, {Kassin},
  {Martin}, {Phillips}, {Rich}, {Small}, {Willmer}, \& {Wilson}}]{noe07}
{Noeske} K.~G. {et~al.}, 2007, \apjl, 660, L43

\bibitem[{{Omand} {et~al}\mbox{.}(2014){Omand}, {Balogh}, \&
  {Poggianti}}]{oma14}
{Omand} C.~M.~B., {Balogh} M.~L., {Poggianti} B.~M., 2014, \mnras, 440, 843

\bibitem[{{Peng} {et~al}\mbox{.}(2010){Peng}, {Lilly}, {Kova{\v c}},
  {Bolzonella}, {Pozzetti}, {Renzini}, {Zamorani}, {Ilbert}, {Knobel},
  {Iovino}, {Maier}, {Cucciati}, {Tasca}, {Carollo}, {Silverman}, {Kampczyk},
  {de Ravel}, {Sanders}, {Scoville}, {Contini}, {Mainieri}, {Scodeggio},
  {Kneib}, {Le F{\`e}vre}, {Bardelli}, {Bongiorno}, {Caputi}, {Coppa}, {de la
  Torre}, {Franzetti}, {Garilli}, {Lamareille}, {Le Borgne}, {Le Brun},
  {Mignoli}, {Perez Montero}, {Pello}, {Ricciardelli}, {Tanaka}, {Tresse},
  {Vergani}, {Welikala}, {Zucca}, {Oesch}, {Abbas}, {Barnes}, {Bordoloi},
  {Bottini}, {Cappi}, {Cassata}, {Cimatti}, {Fumana}, {Hasinger}, {Koekemoer},
  {Leauthaud}, {Maccagni}, {Marinoni}, {McCracken}, {Memeo}, {Meneux}, {Nair},
  {Porciani}, {Presotto}, \& {Scaramella}}]{pen10a}
{Peng} Y.-J. {et~al.}, 2010, \apj, 721, 193

\bibitem[{{Peng} {et~al}\mbox{.}(2012){Peng}, {Lilly}, {Renzini}, \&
  {Carollo}}]{pen12}
{Peng} Y.-J., {Lilly} S.~J., {Renzini} A., {Carollo} M., 2012, \apj, 757, 4

\bibitem[{{Rodighiero} {et~al}\mbox{.}(2011){Rodighiero}, {Daddi},
  {Baronchelli}, {Cimatti}, {Renzini}, {Aussel}, {Popesso}, {Lutz}, {Andreani},
  {Berta}, {Cava}, {Elbaz}, {Feltre}, {Fontana}, {F{\"o}rster Schreiber},
  {Franceschini}, {Genzel}, {Grazian}, {Gruppioni}, {Ilbert}, {Le Floch},
  {Magdis}, {Magliocchetti}, {Magnelli}, {Maiolino}, {McCracken}, {Nordon},
  {Poglitsch}, {Santini}, {Pozzi}, {Riguccini}, {Tacconi}, {Wuyts}, \&
  {Zamorani}}]{rod11}
{Rodighiero} G. {et~al.}, 2011, \apjl, 739, L40

\bibitem[{{Salim} {et~al}\mbox{.}(2007){Salim}, {Rich}, {Charlot},
  {Brinchmann}, {Johnson}, {Schiminovich}, {Seibert}, {Mallery}, {Heckman},
  {Forster}, {Friedman}, {Martin}, {Morrissey}, {Neff}, {Small}, {Wyder},
  {Bianchi}, {Donas}, {Lee}, {Madore}, {Milliard}, {Szalay}, {Welsh}, \&
  {Yi}}]{sal07}
{Salim} S. {et~al.}, 2007, \apjs, 173, 267

\bibitem[{{Schiminovich} {et~al}\mbox{.}(2007){Schiminovich}, {Wyder},
  {Martin}, {Johnson}, {Salim}, {Seibert}, {Treyer}, {Budav{\'a}ri}, {Hoopes},
  {Zamojski}, {Barlow}, {Forster}, {Friedman}, {Morrissey}, {Neff}, {Small},
  {Bianchi}, {Donas}, {Heckman}, {Lee}, {Madore}, {Milliard}, {Rich}, {Szalay},
  {Welsh}, \& {Yi}}]{sch07b}
{Schiminovich} D. {et~al.}, 2007, \apjs, 173, 315

\bibitem[{{Sheth} {et~al}\mbox{.}(2008){Sheth}, {Elmegreen}, {Elmegreen},
  {Capak}, {Abraham}, {Athanassoula}, {Ellis}, {Mobasher}, {Salvato},
  {Schinnerer}, {Scoville}, {Spalsbury}, {Strubbe}, {Carollo}, {Rich}, \&
  {West}}]{she08a}
{Sheth} K. {et~al.}, 2008, \apj, 675, 1141

\bibitem[{{Skibba} {et~al}\mbox{.}(2009){Skibba}, {Bamford}, {Nichol},
  {Lintott}, {Andreescu}, {Edmondson}, {Murray}, {Raddick}, {Schawinski},
  {Slosar}, {Szalay}, {Thomas}, \& {Vandenberg}}]{ski09}
{Skibba} R.~A. {et~al.}, 2009, \mnras, 399, 966

\bibitem[{{Strauss} {et~al}\mbox{.}(2002){Strauss}, {Weinberg}, {Lupton},
  {Narayanan}, {Annis}, {Bernardi}, {Blanton}, {Burles}, {Connolly},
  {Dalcanton}, {Doi}, {Eisenstein}, {Frieman}, {Fukugita}, {Gunn},
  {Ivezi{\'c}}, {Kent}, {Kim}, {Knapp}, {Kron}, {Munn}, {Newberg}, {Nichol},
  {Okamura}, {Quinn}, {Richmond}, {Schlegel}, {Shimasaku}, {SubbaRao},
  {Szalay}, {Vanden Berk}, {Vogeley}, {Yanny}, {Yasuda}, {York}, \&
  {Zehavi}}]{str02}
{Strauss} M.~A. {et~al.}, 2002, \aj, 124, 1810

\bibitem[{{Taylor}(2005)}]{tay05}
{Taylor} M.~B., 2005, in Astronomical Society of the Pacific Conference Series,
  Vol. 347, Astronomical Data Analysis Software and Systems XIV, {Shopbell} P.,
  {Britton} M., {Ebert} R., eds., p.~29

\bibitem[{{van den Bosch} {et~al}\mbox{.}(2008){van den Bosch}, {Aquino},
  {Yang}, {Mo}, {Pasquali}, {McIntosh}, {Weinmann}, \& {Kang}}]{van08b}
{van den Bosch} F.~C., {Aquino} D., {Yang} X., {Mo} H.~J., {Pasquali} A.,
  {McIntosh} D.~H., {Weinmann} S.~M., {Kang} X., 2008, \mnras, 387, 79

\bibitem[{{Whitaker} {et~al}\mbox{.}(2014){Whitaker}, {Franx}, {Leja}, {van
  Dokkum}, {Henry}, {Skelton}, {Fumagalli}, {Momcheva}, {Brammer}, {Labb{\'e}},
  {Nelson}, \& {Rigby}}]{whi14}
{Whitaker} K.~E. {et~al.}, 2014, \apj, 795, 104

\bibitem[{{Willett} {et~al}\mbox{.}(2013){Willett}, {Lintott}, {Bamford},
  {Masters}, {Simmons}, {Casteels}, {Edmondson}, {Fortson}, {Kaviraj}, {Keel},
  {Melvin}, {Nichol}, {Raddick}, {Schawinski}, {Simpson}, {Skibba}, {Smith}, \&
  {Thomas}}]{wil13}
{Willett} K.~W. {et~al.}, 2013, \mnras, 435, 2835

\bibitem[{{York} {et~al}\mbox{.}(2000){York}, {Adelman}, {Anderson},
  {Anderson}, {Annis}, {Bahcall}, {Bakken}, {Barkhouser}, {Bastian}, {Berman},
  {Boroski}, {Bracker}, {Briegel}, {Briggs}, {Brinkmann}, {Brunner}, {Burles},
  {Carey}, {Carr}, {Castander}, {Chen}, {Colestock}, {Connolly}, {Crocker},
  {Csabai}, {Czarapata}, {Davis}, {Doi}, {Dombeck}, {Eisenstein}, {Ellman},
  {Elms}, {Evans}, {Fan}, {Federwitz}, {Fiscelli}, {Friedman}, {Frieman},
  {Fukugita}, {Gillespie}, {Gunn}, {Gurbani}, {de Haas}, {Haldeman}, {Harris},
  {Hayes}, {Heckman}, {Hennessy}, {Hindsley}, {Holm}, {Holmgren}, {Huang},
  {Hull}, {Husby}, {Ichikawa}, {Ichikawa}, {Ivezi{\'c}}, {Kent}, {Kim},
  {Kinney}, {Klaene}, {Kleinman}, {Kleinman}, {Knapp}, {Korienek}, {Kron},
  {Kunszt}, {Lamb}, {Lee}, {Leger}, {Limmongkol}, {Lindenmeyer}, {Long},
  {Loomis}, {Loveday}, {Lucinio}, {Lupton}, {MacKinnon}, {Mannery}, {Mantsch},
  {Margon}, {McGehee}, {McKay}, {Meiksin}, {Merelli}, {Monet}, {Munn},
  {Narayanan}, {Nash}, {Neilsen}, {Neswold}, {Newberg}, {Nichol}, {Nicinski},
  {Nonino}, {Okada}, {Okamura}, {Ostriker}, {Owen}, {Pauls}, {Peoples},
  {Peterson}, {Petravick}, {Pier}, {Pope}, {Pordes}, {Prosapio},
  {Rechenmacher}, {Quinn}, {Richards}, {Richmond}, {Rivetta}, {Rockosi},
  {Ruthmansdorfer}, {Sandford}, {Schlegel}, {Schneider}, {Sekiguchi}, {Sergey},
  {Shimasaku}, {Siegmund}, {Smee}, {Smith}, {Snedden}, {Stone}, {Stoughton},
  {Strauss}, {Stubbs}, {SubbaRao}, {Szalay}, {Szapudi}, {Szokoly}, {Thakar},
  {Tremonti}, {Tucker}, {Uomoto}, {Vanden Berk}, {Vogeley}, {Waddell}, {Wang},
  {Watanabe}, {Weinberg}, {Yanny}, {Yasuda}, \& {SDSS Collaboration}}]{yor00}
{York} D.~G. {et~al.}, 2000, \aj, 120, 1579

\end{thebibliography}

\end{document}